\documentclass[12pt]{article}%
\usepackage{amsfonts}
\usepackage{graphics}
\usepackage{amsmath}
\usepackage{amssymb}
\usepackage{graphicx}%
\providecommand{\U}[1]{\protect\rule{.1in}{.1in}}
\begin{document}
\centerline{\textbf{\Large}}

\vskip 0.8truecm
\centerline{\textbf{\Large Hyperfine Structure and Zeeman Splitting}}
\vskip 0.3truecm
\centerline{\textbf{\Large in Two-Fermion Bound-State Systems}}
\vskip 0.6truecm \centerline{\large Andrei G. Terekidi$^{a }$,
Jurij W. Darewych$^{b }$, Marko Horbatsch$^{c}$}

\vskip 0.5truecm \centerline{\footnotesize \emph{Department of Physics and
Astronomy, York University, Toronto, Ontario, M3J 1P3, Canada}}
\centerline{\footnotesize \emph{$^{a }$terekidi@yorku.ca,
$^{b}$darewych@yorku.ca,$^{c}$marko@yorku.ca } }

\vskip1.6truecm

\begin{center}
\noindent\textbf{Abstract}
\end{center}

A relativistic wave equation for bound states of two fermions with arbitrary
masses which are exposed to a magnetic field is derived from quantum
electrodynamics. The interaction kernels are based upon the generalized
invariant $\widetilde{\mathcal{M}}\mathcal{\,}$-matrices for inter-fermion and
fermion-field interactions. As an application we calculate the energy
corrections in a weak homogeneous $\mathbf{B}$\ field to obtain the Zeeman
splitting of the hyperfine structure (HFS) and $g$-factors in the lowest order
(\textit{i.e.} to $O\left(  \alpha^{4}\right)  )$. Land\'{e} $g$-factors are
presented for several of the first excited states of hydrogen, muonium, and muonic-hydrogen.

\vskip0.6truecm

\noindent\textbf{{\large 1. Introduction}}

\vskip0.4truecm

The relativistic treatment of energy levels of two-fermion atomic systems
(including atomic hydrogen, hydrogen-like ions, helium-3 ion, muonium,
muonic-hydrogen), as well as their fine structure (FS) and hyperfine structure
(HFS) in an external uniform magnetic field (Zeeman effect), is an important
problem. The theoretical knowledge of energy spectra and transition
frequencies provides a test of two-body bound-state QED [1]. One can then
obtain information about the character of the coupling in the system, the
gyromagnetic ratios of the bound particles, the magnetic moments [2], the mass
ratio [2-5], and fundamental physical constants such as the Rydberg constant
$R_{\infty}$, and the fine structure constant $\alpha$ [6]. The Zeeman effect
in the HFS can be used as a diagnostic tool for solar photospheric magnetic
fields [7], fusion research and plasma physics, where the magnetic field is
applied to control the shape and position of the plasma [8].

In the lowest-order approximation the linearly dependent part of the energy
splitting for a two-fermion system placed in a weak static magnetic field
$\mathbf{B}$\ can be written as [1,9-11]%
\begin{equation}
\Delta E_{F,m_{J},j_{1},\ell,s_{1},I}^{ext}=\left(  \mu_{B1}g_{1}+\mu
_{B2}g_{2}\right)  Bm_{F},
\end{equation}
where $F$, $m_{J}$, $j_{1}$, $\ell$, $s_{1}$, $I$\ are quantum numbers, which
characterize the system: $s_{1}$ and $I$ are the spins of the first and second
particle respectively, $\ell$ and $j_{1}$\ represent the orbital and total
angular momentum quantum numbers of the first particle. The total angular
momentum of the system is denoted by the quantum number $F=j_{1}+I,$%
\ $j_{1}+I-1,...,\left\vert j_{1}-I\right\vert $. The projection of the total
angular momentum on the $\mathbf{B}$ direction is $m_{F}=-F,-F+1,...F-1,F$.
The \textquotedblleft Bohr magnetons\textquotedblright for the two particles
are defined as $\displaystyle\mu_{B1}=Q_{1}\hbar/2m_{1}c$, and $\mu
_{B2}=-Q_{2}\hbar/2m_{2}c$, where $Q_{1}$, $Q_{2}>0$). Usually, in our
notation $m_{1}$\ and $m_{2}$\ correspond to the light and heavy particle
respectively.\ Assuming that the energy-level splitting (1) is much smaller
then the HFS splitting, $\Delta E^{ext}<<\Delta E^{HFS}$,\ the Land\'{e}
($g$-) factors $g_{1}$\ and $g_{2}$ take the form [9-11]%
\begin{equation}
g_{1}=g_{j_{1}}\frac{F\left(  F+1\right)  +j_{1}\left(  j_{1}+1\right)
-I\left(  I+1\right)  }{2F\left(  F+1\right)  },
\end{equation}
where%
\begin{equation}
g_{j_{1}}=1+\left(  g_{s_{1}}-1\right)  \frac{j_{1}\left(  j_{1}+1\right)
+s_{1}\left(  s_{1}+1\right)  -\ell\left(  \ell+1\right)  }{2j_{1}\left(
j_{1}+1\right)  },
\end{equation}
and%
\begin{equation}
g_{2}=g_{s_{2}}\frac{F\left(  F+1\right)  -j_{1}\left(  j_{1}+1\right)
+I\left(  I+1\right)  }{2F\left(  F+1\right)  }.
\end{equation}
Here $g_{s_{1}}$ and $g_{s_{2}}$\ are the intrinsic spin magnetic moments of
the constituent particles. According to the Dirac theory a free particle at
rest has $g_{s}=2$. In QED $g_{s}$ is corrected by the anomaly, which to
lowest order is given by the Schwinger correction. For bound particles the
intrinsic moment can be expressed as
\begin{equation}
g_{s_{1,2}}=2+\bigtriangleup g_{s_{1,2}}^{REL}+\bigtriangleup g_{s_{12}}%
^{QED},
\end{equation}
where the terms $\bigtriangleup g^{REL}$, $\bigtriangleup g^{QED}$ represent
the relativistic [9,12,13], and QED corrections respectively (cf. the review
[14]). There is also an additional higher-order contribution to (1),
$\bigtriangleup g_{1,2}^{HFS}\mu_{B1,2}Bm_{F}$, which is caused by the
hyperfine structure (HFS) [15].

The $g$-factors (2) and (4) are not symmetrical, because they were obtained
under the assumption that the orbital motion of the heavy particle can be
neglected. In hydrogen the nucleus contributes a fraction of $m_{1}/\left(
m_{1}+m_{2}\right)  \approx5\times10^{-4}$ to the orbital angular momentum,
while for muonic hydrogen this fraction is $\approx0.1$. The relativistic and
QED corrections in (5) can be comparable with the orbital angular momentum
effects of the heavy particle. Recent high-precision measurements of the
$g$-factor in hydrogen-like systems have reached an accuracy of about
$5\times10^{-9}$ [16,17]. Thus, it is desirable to obtain a more general
result for the $g$-factor in order to overcome the shortcomings of Eqs.~(2,4).
It will be shown that this is particularly important for excited states.

In this work we present an analysis of the HFS of a two-fermion system in an
external magnetic field based upon a reformulation of QED and the variational
Hamiltonian formalism developed earlier [18-20].\ A relativistic two-fermion
wave equation for arbitrary fermion masses is, thus, derived from first
principles. A solution of this equation permits, in principle, to obtain all
QED energy corrections to any order of the coupling constant [18]. In the
present paper we extend the method to derive the integral wave equation in
momentum space for the case where a uniform weak magnetic field is present. We
calculate the Zeeman splitting of the HFS energy levels to $O\left(
\alpha^{4}\right)  $ for all quantum states and unrestricted values for the
fermion masses. We obtain a novel result for the $g$-factor, Eqs.~(38-41), and
demonstrate that it coincides with Eqs.~(2-4) in the case of $m_{2}>>m_{1}$,
as long as the intrinsic moment of $m_{1}$ is restricted to the Dirac value
$g_{s}=2$.

The modification of the wave equations due to the external magnetic field is
presented in Section 2. In Section 3 we provide the classification of the
quantum states, and a partial-wave decomposition of the momentum-space
equations. Section 4 contains expressions for the Zeeman energy splittings of
the HFS levels, and the $g$-factor results. Numerical values for the Land\'{e}
factors are compared with data from Eqs.~(2,4) for various excited states of
hydrogen, muonium and muonic hydrogen. In most expressions we use natural
units $\hbar= c = 1$. \vskip1.6truecm

\noindent\textbf{{\large 2. Bound-State Variational Wave Equation}}

\vskip0.4truecm

For two-fermion systems without external fields wave equations were derived in
[18-19] on the basis of a modified QED Lagrangian [21-22]. With this
Lagrangian a simple Fock-space trial state
\begin{equation}
\left\vert \psi_{trial}\right\rangle =\underset{s_{1}s_{2}}{\sum}\int
d^{3}\mathbf{p}_{1}d^{3}\mathbf{p}_{2}F_{s_{1}s_{2}}(\mathbf{p}_{1}%
,\mathbf{p}_{2})b_{\mathbf{p}_{1}s_{1}}^{\dagger}D_{\mathbf{p}_{2}s_{2}%
}^{\dagger}\left\vert 0\right\rangle ,
\end{equation}
sufficed to obtain HFS levels correct to fourth order in the fine-structure
constant. Here $b_{\mathbf{q}_{1}s_{1}}^{\dagger}$ and $D_{\mathbf{q}_{2}%
s_{2}}^{\dagger}$\ are creation operators for a free fermion of mass $m_{1}$
and an (anti-)fermion of mass $m_{2}$ respectively, and $\left\vert
0\right\rangle $ is the trial vacuum state such that $b_{\mathbf{q}_{1}s_{1}%
}\left\vert 0\right\rangle =D_{\mathbf{q}_{2}s_{2}}\left\vert 0\right\rangle
=0$.

As discussed in section 3 below, the four adjustable functions $F_{s_{1}s_{2}%
}$ must be chosen so that the trial state (17) is an eigenstate of the
relativistic total angular momentum operator, its projection, and parity (as
well as charge conjugation for the case $m_{1}=m_{2}$\ such as positronium).

A variational principle is invoked to obtain a momentum-space wave equation
for the amplitudes [18]:
\begin{align}
0  &  =\sum_{s_{1}s_{2}}\int d^{3}\mathbf{p}_{1}d^{3}\mathbf{p}_{2}\left(
\omega_{p_{1}}+\Omega_{p_{2}}-E\right)  F_{s_{1}s_{2}}(\mathbf{p}%
_{1},\mathbf{p}_{2})\delta F_{s_{1}s_{2}}^{\ast}(\mathbf{p}_{1},\mathbf{p}%
_{2})\\
&  -\frac{m_{1}m_{2}}{\left(  2\pi\right)  ^{3}}\underset{\sigma_{1}\sigma
_{2}s_{1}s_{2}}{\sum}\int\frac{d^{3}\mathbf{p}_{1}d^{3}\mathbf{p}_{2}%
d^{3}\mathbf{q}_{1}d^{3}\mathbf{q}_{2}}{\sqrt{\omega_{p_{1}}\omega_{q_{1}%
}\Omega_{p_{2}}\Omega_{q_{2}}}}\nonumber\\
&  \times F_{\sigma_{1}\sigma_{2}}(\mathbf{q}_{1},\mathbf{q}_{2})\left(
-i\right)  \widetilde{\mathcal{M}}_{s_{1}s_{2}\sigma_{1}\sigma_{2}}\left(
\mathbf{p}_{1},\mathbf{p}_{2},\mathbf{q}_{1}\mathbf{,q}_{2}\right)  \delta
F_{s_{1}s_{2}}^{\ast}(\mathbf{p}_{1},\mathbf{p}_{2}),\nonumber
\end{align}
where $\omega_{p_{1}}^{2}=\mathbf{p}_{1}^{2}+m_{1}^{2}$ and $\Omega_{p_{1}%
}^{2}=\mathbf{p}_{1}^{2}+m_{2}^{2}$. The interaction is governed by the
generalized invariant $\mathcal{M}$-matrix $\widetilde{\mathcal{M}}%
_{s_{1}s_{2}\sigma_{1}\sigma_{2}}\left(  \mathbf{p}_{1},\mathbf{p}%
_{2},\mathbf{q}_{1}\mathbf{,q}_{2}\right)  $.\ It has the form
\begin{equation}
\mathcal{M}_{s_{1}s_{2}\sigma_{1}\sigma_{2}}^{\left(  1\right)  }\left(
\mathbf{p}_{1},\mathbf{p}_{2},\mathbf{q}_{1}\mathbf{,q}_{2}\right)
\equiv\mathcal{M}_{s_{1}s_{2}\sigma_{1}\sigma_{2}}^{ope}\left(  \mathbf{p}%
_{1},\mathbf{p}_{2},\mathbf{q}_{1}\mathbf{,q}_{2}\right)  +\mathcal{M}%
_{s_{1}s_{2}\sigma_{1}\sigma_{2}}^{ext}\left(  \mathbf{p}_{1},\mathbf{p}%
_{2},\mathbf{q}_{1}\mathbf{,q}_{2}\right)  ,
\end{equation}
where $\mathcal{M}_{s_{1}s_{2}\sigma_{1}\sigma_{2}}^{ope}\left(
\mathbf{p}_{1},\mathbf{p}_{2},\mathbf{q}_{1}\mathbf{,q}_{2}\right)  $ is the
usual invariant matrix element, corresponding to the one-photon exchange
Feynman diagram [19-20].

The element $\mathcal{M}_{s_{1}s_{2}\sigma_{1}\sigma_{2}}^{ext}$ represents
the interaction with a given external classical field $A_{\mu}^{ext}$%
\begin{align}
&  \mathcal{M}_{s_{1}s_{2}\sigma_{1}\sigma_{2}}^{ext}\left(  \mathbf{p}%
_{1},\mathbf{p}_{2},\mathbf{q}_{1}\mathbf{,q}_{2}\right) \\
&  =i\left(  2\pi\right)  ^{3/2}\left(
\begin{array}
[c]{c}%
\frac{\sqrt{\Omega_{\mathbf{p}_{2}}\Omega_{q_{2}}}}{m_{2}}A_{\mu}%
^{ext}(\mathbf{p}_{1}-\mathbf{q}_{1})\overline{u}\left(  \mathbf{p}_{1}%
,s_{1}\right)  \left(  -iQ_{1}\right)  \gamma^{\mu}u\left(  \mathbf{q}%
_{1},\sigma_{1}\right)  \delta_{s_{2}\sigma_{2}}\\
+\frac{\sqrt{\omega_{\mathbf{p}_{1}}\omega_{\mathbf{q}_{1}}}}{m_{1}}A_{\mu
}^{ext}(\mathbf{q}_{2}-\mathbf{p}_{2})\overline{V}\left(  \mathbf{p}%
_{2},\sigma_{2}\right)  \left(  -iQ_{2}\right)  \gamma^{\mu}V\left(
\mathbf{q}_{2},s_{2}\right)  \delta_{s_{1}\sigma_{1}}%
\end{array}
\right)  .\nonumber
\end{align}

The Ansatz (6) can not accommodate processes that include the emission or
absorption of real, physical (as opposed to virtual) photons. Such radiative
processes could be included by generalizing the trial state. Here we limit
ourselves to the form (6), \textit{i.e.}, the effects of radiative decay or
absorption of radiation are ignored in the present work.

In order to obtain the Lande factors we evaluate the $\mathcal{M}_{s_{1}%
s_{2}\sigma_{1}\sigma_{2}}^{ext}$ matrix (9) in a stationary uniform magnetic
field $\mathbf{B}=B\mathbf{\hat{z}}$. The vector potential can be chosen as%
\begin{equation}
A_{1}^{ext}\left(  \mathbf{x}\right)  =-\frac{1}{2}yB,\;\;\ \ \;A_{2}%
^{ext}\left(  \mathbf{x}\right)  =\frac{1}{2}xB,\ \ \ \ \ \ \ \ A_{0}%
^{ext}\left(  \mathbf{x}\right)  =A_{3}^{ext}\left(  \mathbf{x}\right)  =0.
\end{equation}
The inverse Fourier transform of the non-zero components yields%
\begin{equation}
A_{1}^{ext}(\mathbf{k})=\frac{\left(  2\pi\right)  ^{3/2}iB}{2}\delta\left(
k_{x}\right)  \frac{d\delta\left(  k_{y}\right)  }{dk_{y}}\delta\left(
k_{z}\right)  ,\ \ \ \ A_{2}^{ext}(\mathbf{k})=\mathbf{-}\frac{\left(
2\pi\right)  ^{3/2}iB}{2}\frac{d\delta\left(  k_{x}\right)  }{dk_{x}}%
\delta\left(  k_{y}\right)  \delta\left(  k_{z}\right)  .
\end{equation}
Using the semi-relativistic expansion%
\begin{align}
\overline{u}\left(  \mathbf{p}_{1},s_{1}\right)  \gamma^{1}u\left(
\mathbf{q}_{1},\sigma_{1}\right)   &  =\frac{1}{2m_{1}c}\varphi_{s_{1}%
}^{\dagger}\left(  i\left[  \overrightarrow{\mathbf{\sigma}}_{1}\times\left(
\mathbf{p}_{1}-\mathbf{q}_{1}\right)  \right]  +\mathbf{q}_{1}+\mathbf{p}%
_{1}\right)  _{1}\varphi_{\sigma_{1}},\\
\overline{u}\left(  \mathbf{p}_{1},s_{1}\right)  \gamma^{2}u\left(
\mathbf{q}_{1},\sigma_{1}\right)   &  =\frac{1}{2m_{1}c}\varphi_{s_{1}%
}^{\dagger}\left(  i\left[  \overrightarrow{\mathbf{\sigma}}_{1}\times\left(
\mathbf{p}_{1}-\mathbf{q}_{1}\right)  \right]  +\mathbf{q}_{1}+\mathbf{p}%
_{1}\right)  _{2}\varphi_{\sigma_{1}},\nonumber
\end{align}
where $\varphi_{1}^{\dagger}=[1\ 0]$, $\varphi_{2}^{\dagger}=[0\ 1]$,and
$\left(  \omega_{p_{1}}\omega_{q_{1}}\right)  ^{1/2}\simeq m_{1}$, and a
similar expansion for anti-particle spinors we obtain%
\begin{align}
&  \mathcal{M}_{s_{1}s_{2}\sigma_{1}\sigma_{2}}^{ext}\left(  \mathbf{p}%
_{1},\mathbf{p}_{2},\mathbf{q}_{1}\mathbf{,q}_{2}\right) \\
&  =\frac{\left(  2\pi\right)  ^{3/2}}{2c}\left(
\begin{array}
[c]{c}%
\frac{Q_{1}}{m_{1}}A_{j}^{ext}(\mathbf{p}_{1}-\mathbf{q}_{1})\varphi_{s_{1}%
}^{\dagger}\left(  i\left[  \overrightarrow{\mathbf{\sigma}}_{1}\times\left(
\mathbf{p}_{1}-\mathbf{q}_{1}\right)  \right]  +\mathbf{q}_{1}+\mathbf{p}%
_{1}\right)  _{j}\varphi_{\sigma_{1}}\delta_{s_{2}\sigma_{2}}\\
+\frac{Q_{2}}{m_{2}}A_{j}^{ext}(\mathbf{q}_{2}-\mathbf{p}_{2})\chi_{\sigma
_{2}}^{\dagger}\left(  i\left[  \overrightarrow{\mathbf{\sigma}}_{2}%
\times\left(  \mathbf{p}_{2}-\mathbf{q}_{2}\right)  \right]  +\mathbf{q}%
_{2}+\mathbf{p}_{2}\right)  _{j}\chi_{s_{2}}\delta_{s_{1}\sigma_{1}}%
\end{array}
\right)  ,\nonumber
\end{align}
where $\chi_{1}^{\dagger}=[0\ 1]$, $\chi_{2}^{\dagger}=-[1\ 0]$, and $j=1,2$.
It is straightforward to show that%
\begin{equation}
\left(  \mathbf{q}_{1}\right)  _{j}\ A_{j}^{ext}(\mathbf{p}_{1}-\mathbf{q}%
_{1})=-\frac{\left(  2\pi\right)  ^{3/2}B}{2}\widehat{L}_{1z}\left(
\mathbf{q}_{1}\right)  \delta\left(  \mathbf{p}_{1}-\mathbf{q}_{1}\right)  ,
\end{equation}
and%
\begin{align}
&  A_{j}^{ext}(\mathbf{p}_{1}-\mathbf{q}_{1})\varphi_{s_{1}}^{\dagger}\left(
i\left[  \overrightarrow{\mathbf{\sigma}}_{1}\times\left(  \mathbf{p}%
_{1}-\mathbf{q}_{1}\right)  \right]  +\mathbf{q}_{1}+\mathbf{p}_{1}\right)
_{j}\varphi_{\sigma_{1}}\\
&  =-\left(  2\pi\right)  ^{3/2}B\left(  \varphi_{s_{1}}^{\dagger}\sigma
_{1z}\varphi_{\sigma_{1}}+\delta_{s_{1}\sigma_{1}}\widehat{L}_{1z}\left(
\mathbf{q}_{1}\right)  \right)  \delta^{3}\left(  \mathbf{p}_{1}%
-\mathbf{q}_{1}\right)  ,\nonumber
\end{align}
where $\widehat{L}_{1z}\left(  \mathbf{q}_{1}\right)  $\ is the $z$-component
of the angular momentum operator of the particle with mass $m_{1}$%
\begin{equation}
\widehat{L}_{1z}\left(  \mathbf{q}_{1}\right)  =-i\left(  q_{1x}\frac
{\partial}{\partial q_{1y}}-q_{1y}\frac{\partial}{\partial q_{1x}}\right)  .
\end{equation}
Taking $\varphi_{s_{1}}$\ to be the eigenstates of the spin operator
$\widehat{S}_{1z}=\frac{1}{2}\widehat{\sigma}_{1z}$, and using a similar
procedure for the second particle,\ we obtain%
\begin{align}
&  \mathcal{M}_{s_{1}s_{2}\sigma_{1}\sigma_{2}}^{ext}\left(  \mathbf{p}%
_{1},\mathbf{p}_{2},\mathbf{q}_{1}\mathbf{,q}_{2}\right) \\
&  =-\frac{\left(  2\pi\right)  ^{3}B}{2c}\left(
\begin{array}
[c]{c}%
\frac{Q_{1}}{m_{1}}\left(  2\varphi_{s_{1}}^{\dagger}\widehat{S}_{1z}%
\varphi_{\sigma_{1}}+\delta_{s_{1}\sigma_{1}}\widehat{L}_{1z}\left(
\mathbf{q}_{1}\right)  \right)  \delta_{s_{2}\sigma_{2}}\delta^{3}\left(
\mathbf{p}_{1}-\mathbf{q}_{1}\right) \\
-\frac{Q_{2}}{m_{2}}\left(  2\chi_{\sigma_{2}}^{\dagger}\widehat{S}_{2z}%
\chi_{s_{2}}+\delta_{\sigma_{2}s_{2}}\widehat{L}_{2z}\left(  \mathbf{q}%
_{2}\right)  \right)  \delta_{s_{1}\sigma_{1}}\delta^{3}\left(  \mathbf{p}%
_{2}-\mathbf{q}_{2}\right)
\end{array}
\right)  ,\nonumber
\end{align}
or%
\begin{align}
&  \mathcal{M}_{s_{1}s_{2}\sigma_{1}\sigma_{2}}^{ext}\left(  \mathbf{p}%
_{1},\mathbf{p}_{2},\mathbf{q}_{1}\mathbf{,q}_{2}\right) \\
&  =-\left(  2\pi\right)  ^{3}B\left(
\begin{array}
[c]{c}%
\mu_{B1}\left(  2\tilde{m}_{\sigma_{1}}+\widehat{L}_{1z}\left(  \mathbf{q}%
_{1}\right)  \right)  \delta^{3}\left(  \mathbf{p}_{1}-\mathbf{q}_{1}\right)
\\
-\mu_{B2}\left(  2\tilde{m}_{\sigma_{2}}+\widehat{L}_{2z}\left(
\mathbf{q}_{2}\right)  \right)  \delta^{3}\left(  \mathbf{p}_{2}%
-\mathbf{q}_{2}\right)
\end{array}
\right)  \delta_{s_{2}\sigma_{2}}\delta_{s_{1}\sigma_{1}},\nonumber
\end{align}
where the spin projection quantum numbers $\tilde{m}_{\sigma}$ can take the
values $\pm1/2$. The quantities $\mu_{B1}$ and $\mu_{B2}$ are the
\textquotedblleft Bohr magnetons" defined in the previous section. As
expected, a unit of spin interacts with a magnetic field twice as strongly as
a unit of orbital angular momentum.

By going to the next order in the expansion of the invariant $\mathcal{M}$
matrix one can obtain self-energy corrections, which lead to divergent loop
integrals that have to be cured by charge renormalization. The vertex term
modifies the Dirac value of the magnetic moment by a factor $\left(
1+k\right)  $, where $k$ is the anomaly (Schwinger correction). This factor
can be included in our calculation by a replacement $2 \tilde m_{\sigma_{1}}%
$\ and $2 \tilde m_{\sigma_{2}}$ in Eq.~(18) by $g_{s_{1}}\tilde m_{\sigma
_{1}}$ and $g_{s_{2}}\tilde m_{\sigma_{2}}$ respectively, where $g_{s_{1,2}%
}/2=1+k_{1,2}$. The anomaly is the lowest-order\ QED correction to the $g$
factor $\bigtriangleup g_{s_{12}}^{QED}=2k_{1,2}$ in Eq. (5).

\vskip0.8truecm

\noindent\textbf{{\large 3. Partial-wave decomposition and radial wave
equations}}

\vskip0.4truecm

The present work is an extension of Ref.~[18], in which the partial-wave
decomposition of the wave equation has been provided. The external magnetic
field is treated as a first-order perturbation which implies that the quantum
labels for the eigenstates do not change. The restrictions on the magnetic
field strength to justify a perturbative treatment of Eq.(18) are%
\begin{equation}
B\lesssim\min\left[  \frac{\alpha^{4}m_{r}c^{2}}{\mu_{B1}},\frac{\alpha
^{4}m_{r}c^{2}}{\mu_{B2}}\right]  ,
\end{equation}
where $\alpha=Q_{1}Q_{2}/4\pi$, and $m_{r}=m_{1}m_{2}/\left(  m_{1}%
+m_{2}\right)  $ is the reduced mass. A more explicit restriction on $B$\ will
be presented in Section 4.

As outlined in Ref.~[18] the trial state (6) is taken to be an eigenstate of
total linear momentum $\widehat{\mathbf{P}}$, total angular momentum squared
$\widehat{\mathbf{J}}^{2}$, its projection $\widehat{J}_{3}$, and parity
$\widehat{\mathcal{P}}$. It is natural to work in the rest frame, where the
total linear momentum vanishes. In this frame the adjustable functions take
the form $F_{s_{1}s_{2}}(\mathbf{p}_{1},\mathbf{p}_{2})=\delta\left(
\mathbf{p}_{1}+\mathbf{p}_{2}\right)  F_{s_{1}s_{2}}(\mathbf{p}_{1})$, where
$F_{s_{1}s_{2}}(\mathbf{p}_{1})$ (using $\mathbf{p}_{1} \equiv\mathbf{p}$) can
be written as
\begin{equation}
F_{s_{1}s_{2}}(\mathbf{p})=\sum_{\ell_{s_{1}s_{2}}}\sum_{m_{s_{1}s_{2}}%
}f_{s_{1}s_{2}}^{\ell_{s_{1}s_{2}}m_{s_{1}s_{2}}}\left(  p\right)
Y_{\ell_{s_{1}s_{2}}}^{m_{s_{1}s_{2}}}(\widehat{\mathbf{p}}),
\end{equation}
and $Y_{\ell_{s_{1}s_{2}}}^{m_{s_{1}s_{2}}}(\widehat{\mathbf{p}})$\ are the
usual spherical harmonics. Here and henceforth we will use the notation
$p=\left\vert \mathbf{p}\right\vert $ etc., while four-vectors will be written
as $p^{\mu}$. The orbital indices $\ell_{s_{1}s_{2}}$and $m_{s_{1}s_{2}}$ and
the radial functions $f_{s_{1}s_{2}}^{\ell_{s_{1}s_{2}}m_{s_{1}s_{2}}}\left(
p\right)  $ depend on the spin variables $s_{1}$ and $s_{2}$. In the rest
frame, the operators $\widehat{L}_{1z}\left(  \mathbf{q}\right)  $\ and
$\widehat{L}_{2z}\left(  \mathbf{q}\right)  $\ can be expressed in terms of
the orbital angular momentum operator, $\widehat{L}_{z}\left(  \mathbf{q}%
\right)  $, of the relative motion:
\begin{equation}
\widehat{L}_{1z}\left(  \mathbf{q}\right)  =\frac{m_{2}}{m_{1}+m_{2}}%
\widehat{L}_{z}\left(  \mathbf{q}\right)  ,\;\;\;\;\;\;\;\;\widehat{L}%
_{2z}\left(  -\mathbf{q}\right)  =\frac{m_{1}}{m_{1}+m_{2}}\widehat{L}%
_{z}\left(  \mathbf{q}\right)  .
\end{equation}

The substitution of the partial-wave expansion (20) into the rest-frame form
of Ansatz (6)
leads to two categories of relations among the adjustable functions
$F_{s_{1}s_{2}}(\mathbf{p})$:

\vskip0.4truecm

{\normalsize \noindent}\textit{(i) The spin-mixed (quasi-singlet and
quasi-triplet) states }

\noindent In this case we have $\ell_{s_{1}s_{2}}\equiv\ell=J$, and the
general solution under the condition of well-defined $\widehat{\mathbf{P}}$,
$\widehat{\mathbf{J}}^{2}$, $\widehat{J}_{3}$, and $\widehat{\mathcal{P}}$ can
be expressed with the help of Dirac $\Gamma$ matrices as [18]
\begin{equation}
F_{s_{1}s_{2}}\left(  \mathbf{p}\right)  =\overline{u}_{\mathbf{p}s_{1}}%
\Gamma_{m_{s_{1}s_{2}}}^{J\left(  sg\right)  }\left(  \widehat{\mathbf{p}%
}\right)  V_{-\mathbf{p}s_{2}}f_{J}^{\left(  sg\right)  }(p)+\overline
{u}_{\mathbf{p}s_{1}}\Gamma_{m_{s_{1}s_{2}}}^{J\left(  tr\right)  }\left(
\widehat{\mathbf{p}}\right)  V_{-\mathbf{p}s_{2}}f_{J}^{\left(  tr\right)
}(p).
\end{equation}
Here $f_{J}^{\left(  sg\right)  }(p)$ and $f_{J}^{\left(  tr\right)  }(p)$ are
radial functions to be determined. They represent the contributions of
spin-singlet and spin-triplet states, \textit{i.e.}, the total spin is not
conserved in general.

\vskip0.4truecm

{\normalsize \noindent}\textit{ (ii) The }$\ell$-\textit{mixed triplet states}

\noindent These states occur for $\ell_{s_{1}s_{2}}\equiv\ell=J\mp1$. Their
radial decomposition can be written as%
\begin{equation}
F_{s_{1}s_{2}}\left(  \mathbf{p}\right)  =\overline{u}_{\mathbf{p}s_{1}}%
\Gamma_{m_{s_{1}s_{2}}}^{J-1}\left(  \widehat{\mathbf{p}}\right)
V_{-\mathbf{p}s_{2}}f_{J-1}(p)+\overline{u}_{\mathbf{p}s_{1}}\Gamma
_{m_{s_{1}s_{2}}}^{J+1}\left(  \widehat{\mathbf{p}}\right)  V_{-\mathbf{p}%
s_{2}}f_{J+1}(p).
\end{equation}
The system in these states is characterized by $J,$ $m_{J},$ and
$\mathcal{P}=(-1)^{J}$, and $\ell$ is not a good quantum number. The two
radial functions $f_{J-1}(p)$ and $f_{J+1}(p)$ correspond to the cases
$\ell=J-1$ and $\ell=J+1$. Mixing of this type occurs only for principal
quantum number $n \ge3$.

From the variational method we obtain a system of coupled radial equations
expressed in matrix form as
\begin{equation}
\left(  \omega_{p}+\Omega_{p}-E\right)  \mathbb{F}\left(  p\right)
=\frac{m_{1}m_{2}}{\left(  2\pi\right)  ^{3}}\int\frac{q^{2}dq}{\sqrt
{\omega_{p}\omega_{q}\Omega_{p}\Omega_{q}}}\mathbb{K}\left(  p,q\right)
\mathbb{F}\left(  q\right)  ,
\end{equation}
where $\omega_{p}^{2}=\mathbf{p}^{2}+m_{1}^{2}$ and $\Omega_{p}^{2}%
=\mathbf{p}^{2}+m_{2}^{2}$, and $q=\left\vert \mathbf{q}\right\vert $ as
already mentioned. Here $\mathbb{F}\left(  p\right)  $\ and $\mathbb{K}\left(
p,q\right)  $\ are matrices composed of radial functions and kernels
respectively. The kernel matrix $\mathbb{K}=\mathbb{K}^{ope}+\mathbb{K}^{ext}$
is made up of one-photon-exchange and external-field parts. Explicit
expressions for $\mathbb{K}^{ope}$ can be found in Ref.[18], while the
external-field contributions are calculated in this work.

For the spin-mixed states the two-component Fock-space amplitude is given as
\begin{equation}
\mathbb{F}\left(  p\right)  =\left[
\begin{array}
[c]{c}%
f_{J}^{\left(  sg\right)  }(p)\\
f_{J}^{\left(  tr\right)  }(p)
\end{array}
\right]  .
\end{equation}
The equations imply a mixing of spin and radial variables, and the radial
equations are usually coupled. We apply a unitary transformation with rotation
angle $\beta$ to the spin part of function (22) to diagonalize the
kernel-matrix. The diagonalization can be carried out for arbitrary $p$ and
$q$ (cf. Eq.~(55) in the Appendix), and defines a new quasi-spin basis
\begin{equation}
\left\vert s_{1},s_{2},\ell,\widetilde{s},J,m_{J}\right\rangle =C_{1}%
\left\vert s_{1},s_{2},\ell,S=0,J,m_{J}\right\rangle +C_{2}\left\vert
s_{1},s_{2},\ell,S=1,J,m_{J}\right\rangle ,
\end{equation}
where $\ell=J$, $S$ is the total spin of the system, and $\widetilde{s}=0$ for
quasi-singlet and $\widetilde{s}=1$ for quasi-triplet states. The coefficients
used to express the new basis states in terms of the previously defined
singlet and triplet states are found to be $C_{1}=\sqrt{\left(  1+\xi\right)
/2}$, $C_{2}=-\sqrt{\left(  1-\xi\right)  /2}$, for the quasi-singlet states,
and $C_{1}=\sqrt{\left(  1-\xi\right)  /2}$, $C_{2}=\sqrt{\left(
1+\xi\right)  /2}$ for the quasi-triplet states. Here the rotation angle
$\beta$ has been replaced for convenience according to $\tan{2\beta}%
=\sqrt{1-\xi^{2}}/\xi$.

The quasi-singlet and quasi-triplet states are both characterized by the same
quantum numbers $J$, $m_{J}$ and $\mathcal{P}=(-1)^{J+1}$, and they mix the
states given in the $LS$\ coupling representation. The states are labeled for
convenience not by the quasi-spin $z$-projection $t_{3}=\mp1/2$, but rather by
$\widetilde{s}=t_{3}+1/2$, which takes on the values of $0,1$. In the Appendix
the kernels for spin-mixed states are given explicitly in order to solve for
the angle $\beta$, \textit{i.e.}, to determine the $\xi$-values.

In the limit $m_{2}>>m_{1}$ the total angular momenta of the first and the
second particles are $j_{1}=\ell_{1}\pm1/2$, $j_{2}=s_{2}=1/2$, where
$\ell_{1}=\ell$. In this case $j_{1}$ can be used as a good quantum number,
and the role of the indices $\tilde s_{s}$, $\tilde s_{t}$ are played by
$j_{1}=\ell_{1}+1/2$ and $j_{1}=\ell_{1}-1/2$ respectively. In this case the
coefficients $C_{1}$ and $C_{2}$ reduce to C-G coefficients
\begin{equation}
C_{1,2}=\left(  -1\right)  ^{1/2+1/2+\ell_{1}+j_{1}}\sqrt{\left(  2S+1\right)
\left(  2j_{1}+1\right)  }\left\{
\begin{array}
[c]{ccc}%
1/2 & 1/2 & S\\
\ell_{1} & \ell_{1} & j_{1}%
\end{array}
\right\}  .
\end{equation}
Note that the one-body limit corresponds to the $j_{1}j_{2}$ coupling
representation, which can not be used in the general case of arbitrary masses
since $j_{1}$ and $j_{2}$ are not independent (they are related through the
common angular momentum $\ell$). For positronium the quasi-states become true
singlet ($C_{2}=0$) and triplet ($C_{1}=0$) states with different charge
conjugation quantum numbers.

We now proceed to calculate the kernels $\mathcal{K}_{mn}^{ext}\left(
p,q\right)  $ associated with the classical external field $A_{\mu}^{ext}$.
\ Using Eq.~(9) for $\mathcal{M}_{s_{1}s_{2}\sigma_{1}\sigma_{2}}^{ext}$ taken
in the rest frame, we obtain%
\begin{align}
&  \mathcal{K}_{mn}^{ext}\left(  p,q\right)  =-\frac{\left(  \pi/2\right)
^{3/2}}{N\left(  m_{1}m_{2}\right)  ^{2}}\int d^{3}\widehat{\mathbf{p}}%
d^{3}\widehat{\mathbf{q}}\\
&  \times Tr\left(
\begin{array}
[c]{c}%
Q_{1}\sqrt{\Omega_{q}\Omega_{q}}A_{\mu}^{ext}(\mathbf{p}-\mathbf{q})\left(
\gamma^{\lambda}q_{\lambda}+m_{1}\right)  \gamma^{\mu}\left(  \gamma^{\lambda
}q_{\lambda}+m_{1}\right)  \Gamma^{n}\left(  \widehat{\mathbf{q}}\right)
\left(  \gamma^{\lambda}\widetilde{q}_{\lambda}-m_{2}\right)  \Gamma^{\prime
m}\left(  \widehat{\mathbf{p}}\right) \\
-Q_{2}\sqrt{\omega_{p}\omega_{p}}A_{\mu}^{ext}(\mathbf{q}-\mathbf{p})\left(
\gamma^{\lambda}q_{\lambda}+m_{1}\right)  \Gamma^{n}\left(  \widehat
{\mathbf{q}}\right)  \left(  \gamma^{\lambda}\widetilde{q}_{\lambda}%
-m_{2}\right)  \gamma^{\mu}\left(  \gamma^{\lambda}\widetilde{q}_{\lambda
}-m_{2}\right)  \Gamma^{\prime m}\left(  \widehat{\mathbf{p}}\right)
\end{array}
\right)  ,\nonumber
\end{align}
where $q=\left(  \omega_{p},\mathbf{q}\right)  $, and $\widetilde{q}=\left(
\Omega_{q},-\mathbf{q}\right)  $. The $\Gamma$\ -matrices correspond to the
various $J^{\mathcal{P}}$\ states. The evaluation of these kernels would allow
one to obtain all relativistic corrections to the $g$-factor (5), however this
is a formidable task. To determine the lowest-order effect it is sufficient to
use the nonrelativistic limit ($q^{2}/m^{2}<<1$). In this case the kernels
(28) take the form%
\begin{align}
&  \mathcal{K}_{mn}^{ext}\left(  p,q\right)  =-\frac{\left(  \pi/2\right)
^{3/2}}{N}\int d^{3}\widehat{\mathbf{p}}d^{3}\widehat{\mathbf{q}}\\
&  \times Tr\left(
\begin{array}
[c]{c}%
Q_{1}A_{\mu}^{ext}(\mathbf{p}-\mathbf{q})\left(  \gamma^{0}+I\right)
\gamma^{\mu}\left(  \gamma^{0}+I\right)  \Gamma^{n}\left(  \widehat
{\mathbf{q}}\right)  \left(  \gamma^{0}-I\right)  \Gamma^{\prime m}\left(
\widehat{\mathbf{p}}\right) \\
-Q_{2}A_{\mu}^{ext}(\mathbf{q}-\mathbf{p})\left(  \gamma^{0}+I\right)
\Gamma^{n}\left(  \widehat{\mathbf{q}}\right)  \left(  \gamma^{0}-I\right)
\gamma^{\mu}\left(  \gamma^{0}-I\right)  \Gamma^{\prime m}\left(
\widehat{\mathbf{p}}\right)
\end{array}
\right)  .\nonumber
\end{align}
These are evaluated for a stationary uniform magnetic field (10). The results
are given separately for the two types of states:

{\normalsize \noindent}\textit{(i) The spin-mixed states} ($\ell
=J,\;\ J\geq1,\;\mathcal{P}=(-1)^{J+1}$)

\noindent In contrast to $\mathbb{K}^{\left(  ope\right)  }\left(  p,q\right)
$ the kernel matrix $\mathbb{K}^{\left(  ext\right)  }\left(  p,q\right)  $ is
not diagonal in the basis of the quasi-singlet $\left\vert sg_{q}\right\rangle
$ and quasi-triplet $\left\vert tr_{q}\right\rangle $ states, and can be
written as
\begin{align}
&  \mathcal{K}_{11}^{\left(  ext\right)  }\left(  p,q\right) \\
&  =-\frac{\left(  2\pi\right)  ^{3}}{2c}\left(
\begin{array}
[c]{c}%
\frac{Q_{1}}{m_{1}}\left(  \left(  1-\frac{1-\xi}{2J\left(  J+1\right)
}\right)  \frac{m_{2}}{m_{1}+m_{2}}+\frac{g_{s_{1}}}{2}\left(  \frac{1-\xi
}{2J\left(  J+1\right)  }-2\frac{\left\vert m_{1}-m_{2}\right\vert }%
{m_{1}+m_{2}}\xi\right)  \right) \\
-\frac{Q_{2}}{m_{2}}\left(  \left(  1-\frac{1-\xi}{2J\left(  J+1\right)
}\right)  \frac{m_{1}}{m_{1}+m_{2}}+\frac{g_{s_{2}}}{2}\left(  \frac{1-\xi
}{2J\left(  J+1\right)  }-2\frac{\left\vert m_{1}-m_{2}\right\vert }%
{m_{1}+m_{2}}\xi\right)  \right)
\end{array}
\right)  Bm_{J},\nonumber
\end{align}%
\begin{align}
&  \mathcal{K}_{22}^{\left(  ext\right)  }\left(  p,q\right) \\
&  =-\frac{\left(  2\pi\right)  ^{3}}{2c}\left(
\begin{array}
[c]{c}%
\frac{Q_{1}}{2m_{1}c}\left(  \left(  1-\frac{1+\xi}{2J\left(  J+1\right)
}\right)  \frac{m_{2}}{m_{1}+m_{2}}+\frac{g_{s_{1}}}{2}\left(  \frac{1+\xi
}{2J\left(  J+1\right)  }+2\frac{\left\vert m_{1}-m_{2}\right\vert }%
{m_{1}+m_{2}}\xi\right)  \right) \\
-\frac{Q_{2}}{2m_{2}c}\left(  \left(  1-\frac{1+\xi}{2J\left(  J+1\right)
}\right)  \frac{m_{1}}{m_{1}+m_{2}}+\frac{g_{s_{2}}}{2}\left(  \frac{1+\xi
}{2J\left(  J+1\right)  }+2\frac{\left\vert m_{1}-m_{2}\right\vert }%
{m_{1}+m_{2}}\xi\right)  \right)
\end{array}
\right)  Bm_{J},\nonumber
\end{align}%
\begin{align}
\mathcal{K}_{12}^{\left(  ext\right)  }\left(  p,q\right)   &  =\mathcal{K}%
_{21}^{\left(  ext\right)  }\left(  p,q\right) \\
&  =-\frac{\left(  2\pi\right)  ^{3}}{2c}\left(
\begin{array}
[c]{c}%
\frac{Q_{1}}{m_{1}}\left(  \frac{\xi}{\sqrt{J\left(  J+1\right)  }}%
\frac{g_{s_{1}}}{2}+2\left(  \frac{m_{1}-m_{2}}{m_{1}+m_{2}}\right)  ^{2}%
\xi^{2}\left(  1-\frac{g_{s_{1}}}{2}\right)  \right) \\
-\frac{Q_{2}}{m_{2}}\left(  \frac{\xi}{\sqrt{J\left(  J+1\right)  }}%
\frac{g_{s_{2}}}{2}+2\left(  \frac{m_{1}-m_{2}}{m_{1}+m_{2}}\right)  ^{2}%
\xi^{2}\left(  1-\frac{g_{s_{2}}}{2}\right)  \right)
\end{array}
\right)  Bm_{J}.\nonumber
\end{align}
Thus, it couples the system (24).

\vskip0.2truecm

{\normalsize \noindent}\textit{(ii) The pure triplet and }$\ell$\textit{-mixed
states} ($\ell=J\mp1,\;\ J\geq1,\;\mathcal{P}=(-1)^{J}$)

\noindent The system (24) can not be decoupled for these states, and the
matrix $\mathbb{K}^{\left(  ope\right)  }\left(  p,q\right)  $\ is not
diagonal [19]. The magnetic part of the kernel is, however, diagonal%
\begin{equation}
\mathbb{K}^{\left(  ext\right)  }\left(  p,q\right)  =-\frac{\left(
2\pi\right)  ^{3}}{2c}\left(  \frac{Q_{1}}{m_{1}}-\frac{Q_{2}}{m_{2}}\right)
\left[
\begin{array}
[c]{cc}%
1 & 0\\
0 & 1
\end{array}
\right]  Bm_{J}.
\end{equation}
All kernels $\mathbb{K}^{\left(  ext\right)  }$ vanish in the case of equal
masses and opposite charges ($Q_{1}=Q_{2}$), as occurs in the positronium
case, where magnetic effects appear only in $O\left(  B^{2}\right)  $ [23].

\vskip0.8truecm

\noindent\textbf{{\large 4. HFS to }}$O\left(  \alpha^{4}\right)  $
\textbf{{\large order\ in a magnetic field }}

\vskip0.4truecm

To obtain results for energy levels to $O\left(  \alpha^{4}\right)  $ we solve
the radial equations (24) perturbatively using hydrogen-like radial functions
(non-relativistic Schr\"{o}dinger form $f_{n,J,m_{J}}^{Sch}\left(  p\right)
$) in momentum space [9]. The energy eigenvalues can be calculated from the
matrix equation, which follows from (24)%
\begin{align}
E\int p^{2}dp\mathbb{F}^{\dagger}\left(  p\right)  \mathbb{F}\left(  p\right)
&  =\int p^{2}dp\left(  \omega_{p}+\Omega_{p}\right)  \mathbb{F}^{\dagger
}\left(  p\right)  \mathbb{F}\left(  p\right) \\
&  -\frac{m_{1}m_{2}}{\left(  2\pi\right)  ^{3}}\int_{0}^{\infty}\frac
{p^{2}dp}{\sqrt{\omega_{p}\Omega_{p}}}\int_{0}^{\infty}\frac{q^{2}dq}%
{\sqrt{\omega_{q}\Omega_{q}}}\mathbb{F}^{\dagger}\left(  p\right)
\mathbb{K}\left(  p,q\right)  \mathbb{F}\left(  q\right)  ,\nonumber
\end{align}
If the system (24) has been decoupled, or the contribution of nondiagonal
elements of the $\mathbb{K}\left(  p,q\right)  $\ matrix with given radial
functions in (34) is zero, Eq.~(34) immediately gives the perturbative
solution for the energy levels. As shown in Ref.~[19], the contribution of the
nondiagonal elements $\mathcal{K}_{12}^{ope}$ and $\mathcal{K}_{21}^{ope}$ in
Eq.~(34) to order $O\left(  \alpha^{4}\right)  $ is zero for the $\ell$-mixing
states. Thus, in the present scheme the energy corrections for $\ell$-mixing
states can be calculated independently for $\ell=J-1$ and $\ell=J+1$ states.
As a result, all triplet states with $\ell=J\mp1$\ can be treated as pure
states. In the case of spin-mixed states the kernel matrix $\mathbb{K}^{ope}%
$\ has been diagonalized in the basis of quasi-states (26), however the
magnetic part of the interaction gives rise to the non-diagonal terms (32).
Since we are solving the system (34) perturbatively, we can use a new basis
$\left\vert ext\right\rangle =C_{1}^{\prime}\left\vert sg_{q}\right\rangle
+C_{2}^{\prime}\left\vert tr_{q}\right\rangle $\, which mixes the quasi-states
with arbitrary constants $C_{1}^{\prime}$\ and $C_{2}^{\prime}$. This leads to
a two-level problem with the solution $E_{n,J,m_{J}}=\left(  H_{11}%
+H_{22}\right)  /2\pm\left(  \left(  \left(  H_{11}-H_{22}\right)  /2\right)
^{2}+H_{12}H_{21}\right)  ^{1/2}$, where $H_{11}=H_{11}^{ope}+H_{11}^{ext}$,
$H_{22}=H_{22}^{ope}+H_{22}^{ext}$, $H_{12}=H_{21}=H_{12}^{ext}=H_{21}^{ext}$.
In our case $\left\vert H_{11}-H_{22}\right\vert >>H_{12}H_{21}$, because the
difference $\left\vert H_{11}-H_{22}\right\vert $\ is of the order of the fine
structure which dominates over the hyperfine splitting and the magnetic
perturbation $H_{12}$. Therefore, we can approximate $E_{n,J,m_{J}}\approx
H_{11},\ H_{22}$.

The results are presented in the form%
\begin{equation}
\Delta E_{n,J,m_{J}}=E_{n,J,m_{J}}-\left(  m_{1}+m_{2}\right)  +\frac{\left(
Z\alpha\right)  ^{2}m_{r}}{2n^{2}}=\Delta E_{n,J}\left(  \alpha^{4}\right)
+\Delta E_{J,m_{J}}^{ext},
\end{equation}
where $Q_{2}=Z Q_{1}$. The energy corrections $\Delta E_{n,J}\left(
\alpha^{4}\right)  $ due to the kernels $\mathbb{K}^{\left(  ope\right)
}\left(  p,q\right)  $\ were obtained previously [19]. The corrections $\Delta
E_{n,J}\left(  \alpha^{4}\right)  $\ \ contain spin-spin interactions that
lead to the HFS which is illustrated in Fig.~1 for the low-lying excited
states. A detailed analysis of the HFS to $O\left(  \alpha^{4}\right)  $ is
provided in [19]. We note that the HFS of the $1S_{1/2}$\ and $2S_{1/2}%
$\ states is obtained in agreement with the known Fermi splittings [9],
\textit{i.e.}, $\Delta E_{HFS}\left(  1S_{1/2}\right)  =\left(  Z\alpha
\right)  ^{4}m_{r}\left(  8m_{r}/3M\right)  $, and $\Delta E_{HFS}\left(
2S_{1/2}\right)  = \left(  Z\alpha\right)  ^{4}m_{r}\left(  m_{r}/3M\right)
$, where $M=m_{1}+m_{2}$. The HFS of states with $\ell>0$, however, is more
complicated [19]. In standard spectroscopic notation it has the form
\begin{align}
\Delta E_{HFS}\left(  n,\ell,s_{s}\right)   &  \equiv\Delta E_{n,J=\ell
+1}-\Delta E_{n,J=\ell,s_{s}}\\
&  =\frac{\left(  Z\alpha\right)  ^{4}m_{r}}{n^{3}}\frac{1}{2\ell+1}\left(
\frac{2\ell+1-\xi^{-1}}{4\ell\left(  \ell+1\right)  }+\frac{2m_{r}}{M}\frac
{1}{2\ell+3}\right)  ,\nonumber
\end{align}%
\begin{align}
\Delta E_{HFS}\left(  n,\ell,s_{t}\right)   &  \equiv\Delta E_{n,J=\ell,s_{t}%
}-\Delta E_{n,J=\ell-1}\\
&  =\frac{\left(  Z\alpha\right)  ^{4}m_{r}}{n^{3}}\frac{1}{2\ell+1}\left(
\frac{2\ell+1-\xi^{-1}}{4\ell\left(  \ell+1\right)  }+\frac{2m_{r}}{M}\frac
{1}{2\ell-1}\right)  ,\nonumber
\end{align}
where the quantity $\xi$\ is defined by Eq.~(56), but with the quantum number
$J$\ replaced by $\ell$. The formulae (36) and (37) are valid for all quantum
numbers $n$, $\ell$\ and for any mass values $m_{1}, m_{2}$. The weak external
field further splits the energy levels. Eqs. (36) and (37) give excellent
agreement with experiment for the HFS [19].

The energy corrections $\Delta E_{J,m_{J}}^{ext}$\ remove the degeneracy with
respect to the $m_{J}$ quantum number. The solution of Eq.~(34) in the
above-made approximation can be written in the form of Eq.~(1) for all states.

For all \textit{pure states} ($\ell=J\mp1$) we obtain the following results:

\noindent for $\ell=J-1$:%
\begin{equation}
g_{1,2}=1-\frac{m_{1,2}}{m_{1}+m_{2}}\frac{J-1}{J}+\left(  \frac{g_{s_{1,2}}%
}{2}-1\right)  \frac{1}{J},
\end{equation}
for $\ell=J+1$:%
\begin{equation}
g_{1,2}=1-\frac{m_{1,2}}{m_{1}+m_{2}}\frac{J+2}{J+1}-\left(  \frac{g_{s_{1,2}%
}}{2}-1\right)  \frac{1}{J+1} .
\end{equation}

For \textit{spin--mixed} states $\ell=J\neq0$ the solution of Eq.~(34), as
mentioned, reduces to a standard two-energy level problem. The diagonal
elements of the kernel matrix give the first-order Zeeman splitting (in
$O\left(  B\right)  $) in the quasi-spin representation (26), which was used
to derive the HFS energies (36) and (37). Note that the non-diagonal elements
give a contribution to higher-order Zeeman splitting corrections.

To first order in the magnetic field strength we obtain the Land\'{e} factors
to be%
\begin{equation}
g_{1}=\frac{m_{2}}{m_{1}+m_{2}}\left(  1-\frac{1\pm\xi}{2J\left(  J+1\right)
}\right)  +\frac{g_{s_{1}}}{2}\left(  \frac{1\pm\xi}{2J\left(  J+1\right)
}\pm2\frac{\left\vert m_{1}-m_{2}\right\vert }{m_{1}+m_{2}}\xi\right)  ,
\end{equation}%
\begin{equation}
g_{2}=\frac{m_{1}}{m_{1}+m_{2}}\left(  1-\frac{1\mp\xi}{2J\left(  J+1\right)
}\right)  +\frac{g_{s_{2}}}{2}\left(  \frac{1\mp\xi}{2J\left(  J+1\right)
}\mp2\frac{\left\vert m_{1}-m_{2}\right\vert }{m_{1}+m_{2}}\xi\right)  ,
\end{equation}
where the upper sign is taken for $sg_{q}$ and lower sign for $tr_{q}$ states
respectively. Our expressions (40-41) are symmetrical with respect to the
masses of the two particles. Obviously all these first-order Zeeman
corrections, $\Delta E_{J,m_{J}}^{ext}$, vanish for the positronium case
($m_{1}=m_{2}=m_{e},$ $Z=1$), as expected. The intrinsic factors $g_{s_{1,2}}%
$\ associated with the spins of the individual particles can include QED corrections.

In the case when $m_{2}>>m_{1}$\ our general results agree with the result
from Eqs.~(2,4) in which the orbital motion of the heavy particle is ignored.
It is only in this limit (as discussed below Eq.~(27)), that the total angular
momenta of the individual particles are not related through the common angular
momentum $\ell$, and can be written as $j_{1}=\ell\pm1/2$, and $j_{2}=1/2$. In
$j_{1}$-$j_{2}$\ coupling, the eigenstates are taken to be the eigenstates of
the operators $\widehat{\mathbf{j}}_{1}^{2}=\left(  \widehat{\mathbf{L}%
}+\widehat{\mathbf{s}}_{1}\right)  ^{2}$, $\widehat{\mathbf{j}}_{2}%
^{2}=\widehat{\mathbf{s}}_{2}^{2}$, $\widehat{\mathbf{J}}^{2}$, and
$\widehat{J}_{z}$, and are designated as $\left\vert j_{1}j_{2}Jm_{J}%
\right\rangle $ in contrast to the spin-mixed $\left\vert LsJm_{J}%
\right\rangle $ and pure states $\left\vert LSJm_{J}\right\rangle $ which
diagonalize the expectation value of the Hamiltonian to order $O\left(
\alpha^{4}\right)  $. To facilitate the comparison we make the following
replacement of quantum numbers: $F\rightarrow J$, $J\rightarrow j_{1}$,
$L\rightarrow\ell_{1}=\ell$, $S\rightarrow s_{1}$, $I\rightarrow s_{2}$. It
follows that for all pure states $\ell=J\mp1$, formulae (38-39) and (2-4) give
the same result, namely,
\begin{equation}
g_{1}=1+\left(  \frac{g_{s_{1}}}{2}-1\right)  \frac{1}{J}%
,\ \ \ \ \ \ \ \ g_{2}=\frac{g_{s_{2}}}{2}\frac{1}{J},
\end{equation}
for $\ell=j_{1}-1/2\overset{or}{=}J-1$ and
\begin{equation}
g_{1}=1-\left(  \frac{g_{s_{1}}}{2}-1\right)  \frac{1}{J+1}%
,\ \ \ \ \ \ \ \ \ g_{2}=-\frac{g_{s_{2}}}{2}\frac{1}{J+1}%
\end{equation}
for $\ell=j_{1}+1/2\overset{or}{=}J+1$.

In the limit $m_{2}>>m_{1}$ the energy levels of spin-mixed states $\Delta
E_{J,m_{J}}^{ext\left(  sg_{q}\right)  }$ and $\Delta E_{J,m_{J}}^{ext\left(
tr_{q}\right)  }$\ reduce to $\Delta E_{j_{1}=\ell+1/2,J,m_{J}}^{ext}$ and
$\Delta E_{j_{1}=\ell-1/2,J,m_{J}}^{ext}$ respectively, and the Land\'{e}
factors given by (40-41) take the form
\begin{equation}
g_{1}=\frac{2J+3}{2J+1}+\left(  \frac{g_{s_{1}}}{2}-1\right)  \frac{1}%
{J},\ \ \ \ g_{2}=-\frac{1}{J+1}-\left(  \frac{g_{s_{2}}}{2}-1\right)
\frac{1}{J+1},
\end{equation}
and for $\ell=j_{1}+1/2\overset{or}{=}J\ $($tr_{q}$)%
\begin{equation}
g_{1}=\frac{2J-1}{2J+1}-\left(  \frac{g_{s_{1}}}{2}-1\right)  \frac{1}%
{J+1},\ \ \ \ \ \ g_{2}=\frac{1}{J}+\left(  \frac{g_{s_{2}}}{2}-1\right)
\frac{1}{J}%
\end{equation}
for $\ell=j_{1}-1/2\overset{or}{=}J\ $($sg_{q}$). Here the decoupling angle
$\beta$ is given by $\xi\approx1/\left(  2\ell+1\right)  $ in the
$m_{2}>>m_{1}$\ case.

Formula (4) gives a similar result for the second particle, but for the
lighter particle Eq.~(2) yields%
\begin{equation}
g_{1}=\frac{2J+3}{2J+1}+\left(  \frac{g_{s_{1}}}{2}-1\right)  \frac
{2J+3}{\left(  2J+1\right)  \left(  J+1\right)  }%
\end{equation}
for ($sg_{q}$) states, and%
\begin{equation}
g_{1}=\frac{2J-1}{2J+1}-\left(  \frac{g_{s_{1}}}{2}-1\right)  \frac
{2J-1}{J\left(  2J+1\right)  }%
\end{equation}
for ($tr_{q}$) states. This result agrees with (44-45) only in the particular
case of $g_{s_{1}}=2$. Note that most theoretical and experimental results are
concerned with $nS_{1/2}\left(  J=1\right)  $ states for which the
\textquotedblleft mass ratio\textquotedblright correction in (38) disappears.
Thus our results will be most useful for $\ell>0$ states.

In Tables 1-3 we present results of our calculations of the $g$-factors for
the first excited states in hydrogen, muonium, and muonic hydrogen
respectively. Only states with non-zero total angular momentum are included.
Eqs.~(40-41) are used for the spin-mixed states $P_{1/2\left(  J=1\right)  }$,
$P_{3/2\left(  J=1\right)  }$, $D_{3/2\left(  J=2\right)  }$, $D_{5/2\left(
J=2\right)  }$, Eq.~(38) is used for the pure state $P_{3/2\left(  J=2\right)
}$.

\vskip0.4truecm

\textbf{Table 1.} $g$-factors for the electron ($g_{1}$) and proton ($g_{2}$)
respectively in excited atomic hydrogen states. Results from the present
calculation, Eqs.~(38,40) for electrons, are compared with Eq.~(2) in the top
half of the table. For protons the bottom half displays the present results
from Eqs.(38,41) in comparison with Eq.~(4). Each row contains in the upper
part the Land\'e factor where the intrinsic $g_{s}$-value is corrected for the
anomaly (see text), while the numbers below are based upon the Dirac value
$g_{s}=2$.

\begin{center}

\vskip0.2truecm

$%
\begin{tabular}
[c]{||c|r|r|r|r|r||}\hline
$pe^{-}$ & $P_{1/2\left(  J=1\right)  }$ & $P_{3/2\left(  J=1\right)  }$ &
$P_{3/2\left(  J=2\right)  }$ & $D_{3/2\left(  J=2\right)  }$ & $D_{5/2\left(
J=2\right)  }$\\\hline
$g_{1}$ using Eqs.~(38), (40) & $%
\begin{array}
[c]{c}%
0.33237\\
0.33296
\end{array}
$ & $%
\begin{array}
[c]{c}%
1.66740\\
1.66622
\end{array}
$ & $%
\begin{array}
[c]{c}%
1.00032\\
0.99973
\end{array}
$ & $%
\begin{array}
[c]{c}%
0.59912\\
0.59951
\end{array}
$ & $%
\begin{array}
[c]{c}%
1.40008\\
1.39949
\end{array}
$\\\hline
$g_{1}$ using Eq.~(2) & $%
\begin{array}
[c]{c}%
0.33294\\
1/3
\end{array}
$ & $%
\begin{array}
[c]{c}%
1.66765\\
5/3
\end{array}
$ & $%
\begin{array}
[c]{c}%
1.00059\\
1
\end{array}
$ & $%
\begin{array}
[c]{c}%
0.59965\\
3/5
\end{array}
$ & $%
\begin{array}
[c]{c}%
1.39945\\
7/5
\end{array}
$\\\hline
$g_{2}$ using Eqs.~(38), (41) & $%
\begin{array}
[c]{c}%
1.79321\\
1.00036
\end{array}
$ & $%
\begin{array}
[c]{c}%
-0.89597\\
-0.49955
\end{array}
$ & $%
\begin{array}
[c]{c}%
0.89670\\
0.50027
\end{array}
$ & $%
\begin{array}
[c]{c}%
0.89691\\
0.50049
\end{array}
$ & $%
\begin{array}
[c]{c}%
-0.59711\\
-0.33283
\end{array}
$\\\hline
$g_{2}$ using Eq.~(4) & $%
\begin{array}
[c]{c}%
1.79285\\
1
\end{array}
$ & $%
\begin{array}
[c]{c}%
-0.89642\\
-1/2
\end{array}
$ & $%
\begin{array}
[c]{c}%
0.89642\\
1/2
\end{array}
$ & $%
\begin{array}
[c]{c}%
0.89642\\
1/2
\end{array}
$ & $%
\begin{array}
[c]{c}%
-0.59762\\
-1/3
\end{array}
$\\\hline
\end{tabular}
\ $
\end{center}

\vskip0.4truecm

\textbf{Table 2.} Same as in Table 1, but for muonium. The Land\'e factor for
the electron is $g_{1}$, and for the muon it is $g_{2}$.

\begin{center}

\vskip0.2truecm

$%
\begin{tabular}
[c]{||c|r|r|r|r|r||}\hline
$\mu^{+}e^{-}$ & $P_{1/2\left(  J=1\right)  }$ & $P_{3/2\left(  J=1\right)  }$
& $P_{3/2\left(  J=2\right)  }$ & $D_{3/2\left(  J=2\right)  }$ &
$D_{5/2\left(  J=2\right)  }$\\\hline
$g_{1}$ using Eqs.~(38), (40) & $%
\begin{array}
[c]{c}%
0.329451\\
0.33004
\end{array}
$ & $%
\begin{array}
[c]{c}%
1.66392\\
1.66274
\end{array}
$ & $%
\begin{array}
[c]{c}%
0.99818\\
0.99759
\end{array}
$ & $%
\begin{array}
[c]{c}%
0.59527\\
0.59566
\end{array}
$ & $%
\begin{array}
[c]{c}%
1.39610\\
1.39551
\end{array}
$\\\hline
$g_{1}$ using Eq.~(2) & $%
\begin{array}
[c]{c}%
0.33294\\
1/3
\end{array}
$ & $%
\begin{array}
[c]{c}%
1.66765\\
5/3
\end{array}
$ & $%
\begin{array}
[c]{c}%
1.0006\\
1
\end{array}
$ & $%
\begin{array}
[c]{c}%
0.59965\\
3/5
\end{array}
$ & $%
\begin{array}
[c]{c}%
1.39945\\
7/5
\end{array}
$\\\hline
$g_{2}$ using Eqs.~(38), (41) & $%
\begin{array}
[c]{c}%
1.00434\\
1.00320
\end{array}
$ & $%
\begin{array}
[c]{c}%
-0.49657\\
-0.49598
\end{array}
$ & $%
\begin{array}
[c]{c}%
0.50299\\
0.50241
\end{array}
$ & $%
\begin{array}
[c]{c}%
0.50491\\
0.50433
\end{array}
$ & $%
\begin{array}
[c]{c}%
-0.32923\\
-0.32884
\end{array}
$\\\hline
$g_{2}$ using Eq.~(4) & $%
\begin{array}
[c]{c}%
1.00117\\
1
\end{array}
$ & $%
\begin{array}
[c]{c}%
-0.50058\\
-1/2
\end{array}
$ & $%
\begin{array}
[c]{c}%
0.50058\\
1/2
\end{array}
$ & $%
\begin{array}
[c]{c}%
0.50058\\
1/2
\end{array}
$ & $%
\begin{array}
[c]{c}%
-0.33372\\
-1/3
\end{array}
$\\\hline
\end{tabular}
$
\end{center}

\vskip0.4truecm

\textbf{Table 3.} Same as in Table 1, but for muonic hydrogen. The Land\'e
factor for the muon is $g_{1}$, and for the proton it is $g_{2}$.

\begin{center}

\vskip0.2truecm

$%
\begin{tabular}
[c]{||c|r|r|r|r|r||}\hline
$p^{+}\mu^{-}$ & $P_{1/2\left(  J=1\right)  }$ & $P_{3/2\left(  J=1\right)  }$
& $P_{3/2\left(  J=2\right)  }$ & $D_{3/2\left(  J=2\right)  }$ &
$D_{5/2\left(  J=2\right)  }$\\\hline
$g_{1}$ using Eqs.~(38), (40) & $%
\begin{array}
[c]{c}%
0.26707\\
0.26765
\end{array}
$ & $%
\begin{array}
[c]{c}%
1.58232\\
1.58116
\end{array}
$ & $%
\begin{array}
[c]{c}%
0.95019\\
0.94960
\end{array}
$ & $%
\begin{array}
[c]{c}%
0.50961\\
0.51000
\end{array}
$ & $%
\begin{array}
[c]{c}%
1.30580\\
1.30521
\end{array}
$\\\hline
$g_{1}$ using Eq.~(2) & $%
\begin{array}
[c]{c}%
0.33295\\
1/3
\end{array}
$ & $%
\begin{array}
[c]{c}%
1.66764\\
5/3
\end{array}
$ & $%
\begin{array}
[c]{c}%
1.00058\\
1
\end{array}
$ & $%
\begin{array}
[c]{c}%
0.59965\\
3/5
\end{array}
$ & $%
\begin{array}
[c]{c}%
1.39946\\
7/5
\end{array}
$\\\hline
$g_{2}$ using Eqs.~(38), (41) & $%
\begin{array}
[c]{c}%
1.85425\\
1.06317
\end{array}
$ & $%
\begin{array}
[c]{c}%
-0.80664\\
-0.41198
\end{array}
$ & $%
\begin{array}
[c]{c}%
0.94682\\
0.55040
\end{array}
$ & $%
\begin{array}
[c]{c}%
0.98584\\
0.58982
\end{array}
$ & $%
\begin{array}
[c]{c}%
-0.50225\\
-0.23836
\end{array}
$\\\hline
$g_{2}$ using Eq.~(4) & $%
\begin{array}
[c]{c}%
1.79285\\
1
\end{array}
$ & $%
\begin{array}
[c]{c}%
-0.89642\\
-1/2
\end{array}
$ & $%
\begin{array}
[c]{c}%
0.89642\\
1/2
\end{array}
$ & $%
\begin{array}
[c]{c}%
0.89642\\
1/2
\end{array}
$ & $%
\begin{array}
[c]{c}%
-0.59762\\
-1/3
\end{array}
$\\\hline
\end{tabular}
$
\end{center}

\vskip0.4truecm

Our calculations (given to five digits after the decimal point) are to be
compared with the ($m_{2}\rightarrow\infty$) results (2-4). Upper values for
each $g$-factor have taken into account the following anomalous magnetic
moment values: $g_{e}/2=1.00118$, \ $g_{p}/2=1.792847$ \ \ $g_{\mu
}/2=1.001166$ [1,9,14]. The intrinsic proton anomaly reflects the fact that it
is not a fundamental particle, while in the case of electrons and muons the
lowest-order radiative correction was included. The lower values in each row
were calculated with $g_{s_{1,2}}=2$. We used the following values for the
mass ratios: $m_{p}/m_{e}\approx1836.15267$ and $m_{\mu}/m_{e}\approx
206.76828$ [1,9,14].

For the case of muonium we find that the deviations between the present
results and those obtained from the one-body limit are in the few-percent
range. The muon as the heavier of the two particles acquires a systematically
increased Land\'e factor, while the values are always lowered for the electron.

For muonic hydrogen the effects are more pronounced, and range from 3 to 25 \%
for the states shown in Table 3. Only those results which take the anomalous
magnetic moment of the proton into account should be considered as physically
relevant. The systematics are similar to those shown in Table 2 for muonium,
with the largest decrease in the Land\'{e} factor observed for the muon in the
$P_{1/2(J=1)}$ state (-25 \%), while the largest increase (19 \%) for the
proton $g$-value occurs in the $D_{5/2(J=2)}$ state.

For atomic hydrogen the effect is smallest due to the small $e/p$ mass ratio.
Given that atomic spectroscopy is far more advanced in hydrogen than in muonic
atoms one should not neglect these corrections. For the two above-mentioned
states which are most affected we observe about 0.1 \% deviations in the
electron and proton Land\'e factors respectively.

As mentioned above, our results are applicable only in low magnetic fields,
such that the hyperfine energy splitting exceeds the Zeeman splitting, namely%
\begin{equation}
B<<\frac{\Delta E_{HFS}\left(  n,\ell\right)  }{\mu_{B}^{\ast}g}.
\end{equation}
Thus, formula (48), for $2P_{3/2}$ states, requires that $B<<300\ $gauss for
muonium and $B<<100\ $gauss for hydrogen.

\vskip0.8truecm

\noindent\textbf{{\large 5. Conclusion}}

\vskip0.4truecm

We have used the Hamiltonian variational method in reformulated QED to derive
relativistic stationary-state equations for two-fermion systems in an external
magnetic field. These equations can include interactions to any order of the
coupling constant, at least in principle. The classification of the states
follows naturally from the conserved quantum numbers which appear in the trial
state (6). For given total angular momentum $J$ there are, in general, coupled
equations, both for mixed-spin states, and for triplet mixed-$\ell$ states
(cf. Eq.~(24)). We present explicit forms for the kernels (momentum-space
potentials) for the case of a constant, weak external magnetic field.

We solved the radial equations perturbatively to obtain the Zeeman splitting
of the HFS to order $O\left(  \alpha^{4}\right)  $, and calculated the
$g$-factors for the system of two bound fermions. Our results are applicable
to all states (\textit{i.e.} for all quantum numbers) and any fermion masses.
In the limit $m_{2}>>m_{1}$ our formulae reproduce the well-known $g$-factor
result. For the spin-mixed states, however, Eq.~(2) is found to be not exact
if the intrinsic magnetic moment is different from the Dirac value $g_{s_{1}}=
2$.

\vskip0.8truecm

{\normalsize \noindent{\textbf{{\large Acknowledgment}}} }

{\normalsize \vskip0.4truecm }

The financial support of the Natural Sciences and Engineering Research Council
of Canada is gratefully acknowledged.

\newpage

\vskip0.8truecm

\noindent\textbf{{\large Appendix. One-photon exchange kernels for the
spin-mixed states to order }}${\Large \alpha}^{4}$

\vskip0.4truecm

We use the notation $z=\left(  p^{2}+q^{2}\right)  /2pq$, and $\mathrm{Q}%
_{J}(z)$\ is the Legendre function of the second kind [24]. The contributions
of the various terms to the kernel are as follows ($\ell=J\ (J\geq
1),\;\mathcal{P}=(-1)^{J+1}$):

\noindent\textit{(i)} orbital term%
\begin{align}
\mathcal{K}_{11}^{\left(  orb\right)  }\left(  p,q\right)   &  =\mathcal{K}%
_{22}^{\left(  orb\right)  }\left(  p,q\right)  =\frac{2\pi Q_{1}Q_{2}}%
{pq}\mathrm{Q}_{J}(z)\\
&  +\frac{\pi Q_{1}Q_{2}}{2m_{1}m_{2}}\left(  \left(  \frac{m_{1}}{m_{2}%
}+\frac{m_{2}}{m_{1}}-\left(  J-1\right)  \right)  \left(  \frac{p}{q}%
+\frac{q}{p}\right)  \mathrm{Q}_{J}(z)+2\left(  J+1\right)  \mathrm{Q}%
_{J+1}(z)\right)  ,\nonumber
\end{align}
\textit{(ii)} spin-orbit interaction%
\begin{equation}
\mathcal{K}_{11}^{\left(  s-o\right)  }\left(  p,q\right)  =0,
\end{equation}%
\begin{equation}
\mathcal{K}_{12}^{\left(  s-o\right)  }(p,q)=-\frac{\pi Q_{1}Q_{2}}%
{2m_{1}m_{2}}\left\vert \frac{m_{1}}{m_{2}}-\frac{m_{2}}{m_{1}}\right\vert
\frac{2\sqrt{J\left(  J+1\right)  }}{2J+1}\left(  \mathrm{Q}_{J+1}\left(
z\right)  -\mathrm{Q}_{J-1}\left(  z\right)  \right)  ,
\end{equation}%
\begin{equation}
\mathcal{K}_{22}^{\left(  s-o\right)  }(p,q)=-\frac{\pi Q_{1}Q_{2}}%
{2m_{1}m_{2}}\left(  \frac{m_{1}}{m_{2}}+\frac{m_{2}}{m_{1}}+4\right)
\frac{1}{2J+1}\left(  \mathrm{Q}_{J+1}\left(  z\right)  -\mathrm{Q}%
_{J-1}\left(  z\right)  \right)  ,
\end{equation}
\textit{(iii)} spin-spin interaction%
\[
\mathcal{K}_{11}^{\left(  s-s\right)  }\left(  p,q\right)  =0,
\]%
\begin{equation}
\mathcal{K}_{22}^{\left(  s-s\right)  }(p,q)=\frac{\pi Q_{1}Q_{2}}{m_{1}m_{2}%
}\frac{1}{2J+1}\left(  \mathrm{Q}_{J+1}\left(  z\right)  -\mathrm{Q}%
_{J-1}\left(  z\right)  \right)  .
\end{equation}
The diagonalization condition
\begin{equation}
\tan2\beta\left(  \mathcal{K}_{22}\left(  p,q\right)  -\mathcal{K}_{11}\left(
p,q\right)  \right)  =2\mathcal{K}_{12}\left(  p,q\right)  .
\end{equation}
determines the parameters $\beta$ and $\xi$.:%
\begin{equation}
\tan2\beta=2\left\vert \frac{m_{1}-m_{2}}{m_{1}+m_{2}}\right\vert
\sqrt{J\left(  J+J\right)  },
\end{equation}
and%
\begin{equation}
\xi=\left(  4\left(  \frac{m_{1}-m_{2}}{m_{1}+m_{2}}\right)  ^{2}J\left(
J+1\right)  +1\right)  ^{-1/2}.
\end{equation}
Therefore, we obtain the diagonalized kernels for the quasi-states%
\begin{align}
&  \mathcal{K}^{\left(  sg_{q}\right)  },\mathcal{K}^{\left(  tr_{q}\right)
}\\
&  =\mathcal{K}_{11}^{\left(  orb\right)  }+\frac{\xi\pm1}{\sqrt{1-\xi^{2}}%
}\mathcal{K}_{12}^{\left(  s-o\right)  }\nonumber\\
&  =\frac{2\pi Q_{1}Q_{2}}{pq}\mathrm{Q}_{J}(z)\nonumber\\
&  +\frac{\pi Q_{1}Q_{2}}{2m_{1}m_{2}}\left(  \left(  \frac{m_{1}}{m_{2}%
}+\frac{m_{2}}{m_{1}}-\left(  J-1\right)  \right)  \left(  \frac{p}{q}%
+\frac{q}{p}\right)  \mathrm{Q}_{J}(z)+2\left(  J+1\right)  \mathrm{Q}%
_{J+1}(z)\right) \nonumber\\
&  -\frac{\pi Q_{1}Q_{2}}{2m_{1}m_{2}}\frac{\xi\pm1}{\xi\left(  2J+1\right)
}\left(  \mathrm{Q}_{J+1}\left(  z\right)  -\mathrm{Q}_{J-1}\left(  z\right)
\right)  .\nonumber
\end{align}
\vskip0.2truecm

\newpage

{\normalsize \vskip0.8truecm }

{\normalsize \noindent{\textbf{{\large References}}} }

{\normalsize \vskip0.4truecm }

{\normalsize \enumerate}

1. V. W. Hughes and G. zu Putlitz, in \textit{Quantum Electrodynamics}, edited
by T. Kinoshita (World Scientific, Singapore, 1990), p. 822.

2. W. Liu, M. G. Boshier, S. Dhawan, O. van Dyck, P. Egan, X. Fei, M.
GrossePerdekamp, V. W. Hughes, M. Janousch, K. Jungmann, D. Kawall, F.G.
Mariam, C. Pillai, R. Prigl, G. zuPutlitz, I. Reinhard, W. Schwarz, P. A.
Thompson, and K. A. Woodle, Phys. Rev. Lett. \textbf{82}, 711 (1999).

3. D. E. Casperson, T. W. Crane, A. B. Denison, P.O. Egan, V.W. Hughes, F. G.
Mariam, H. Orth, H. W. Reist, P. A. Souder, R. D. Stambaugh, P. A. Thompson,
and G. zuPutlitz, Phys. Rev. Lett. \textbf{38}, 956 (1977).

4. F. G. Mariam, W. Beer, P.R. Bolton, P. O. Egan, C. J. Gardner, V. W.
Hughes, D. C. Lu, P. A. Souder, H. Orth, J. Vetter, U. Moser, and G.
zuPutlitz, Phys. Rev. Lett. \textbf{49}, 993 (1982).

5. V. Meyer, S. N. Bagayev, P. E. G. Baird, P. Bakule, M. G. Boshier, A.
Breitruck, S. L. Cornish, S. Dychkov, G. H. Eaton, A. Grossman, D. Hubl, V. W.
Hughes, K. Jungmann, I. C. Lane, Y.-W. Liu, D. Lucas, Y. Matyugin, J. Merkel,
G. zuPutlitz, I. Reinhard, P. G. H. Sandars, R. Santra, P. V. Schmidt, C. A.
Scott, W. T. Toner, M. Towrie, K. Trager, C. Wasser, L. Willmann, and V.
Yakhontov, Phys. Rev. Lett. \textbf{84}, 1136 (2000).

6. V. W. Hughes and T. Kinoshita, Rev. Mod. Phys. \textbf{71}, S133 (1999).

7. A. Lopez Ariste, S. Tomczyk, R. Casini, Astrophysical Journal,
\textbf{580}, 519 (2002).

8. R. C. Isler, Plasma Phys. Control. Fusion \textbf{36}, 171 (1994).

9. H. A. Bethe and E. E. Salpeter, \textit{Quantum Mechanics of One- and
Two-Electron Atoms }(Springer, 1957).

10. M. Mizushima, \textit{Quantum Mechanics of Atomic Spectra and Atomic
Structure} (W. A. Benjamin, 1970). p.331.

11. G. K. Woodgate, \textit{Elementary atomic structure} (Clarendon Press,
Oxford, 1980).

12. G. Breit, Nature, 122, 649 (1928).

13. S. A. Zapryagaev, Opt. Spectrosc. \textbf{47}, 9 (1979).

14. S. G. Karshenboim, arXiv:hep-ph/0509010 v1 (2005).

15. D. L. Moskovkin, N. S. Oreshkina, V. M. Shabaev, T. Beier, G. Plunien, W.
Quint, and G. Soff, Phys. Rev. A \textbf{70}, 032105 (2004).

16. N. Hermanspahn, H. H\"{a}ffner, H. J. Kluge, W. Quint, S. Stahl, J
\ Verd\'{u}, and G. Werth, Phys. Rev. Lett. \textbf{84}, 427 (2000).

17. J. Verd\'{u}, S. Djeki\'{c}, H. Haffner, S. Stahl, T. Valenzuela, M.
Vogel, G. Werth, H. J. Kluge, W. Quint, Phys. Rev. Lett. \textbf{92}, 093002 (2004).

18. A. G. Terekidi, J. W. Darewych, M. Horbatsch, arXiv:
hep-th/0604078\ (2006); Can. J. Phys. in press (2007).

19. A. G. Terekidi, J. W. Darewych, Journal of Mathematical Physics
\textbf{46}, 032302 (2005).

20. A. G. Terekidi, J. W. Darewych, Journal of Mathematical Physics
\textbf{45}, 1474 (2004).

21. J. W. Darewych, Annales Fond. L. de Broglie (Paris) \textbf{23}, 15 (1998).

22. J. W. Darewych, in \textit{Causality and Locality in Modern Physics}, G
Hunter et al. (eds.), p. 333, (Kluwer, 1998).

23. V. B. Berestetskii, E. M. Lifshitz, L. P. Pitaevskii, \textit{Relativistic
Quantum Theory} (Pergamon Press, 1971). p.287.

24. G. Arfken and H. Weber, \textit{Mathematical Methods for Physicists}
(Academic Press, 2001), p.805.
\end{document}